# High quality sandwiched black phosphorus heterostructure and its quantum oscillations

Xiaolong Chen[1,†], Yingying Wu[1,†], Zefei Wu[1], Shuigang Xu[1], Lin Wang[2], Yu Han[1], Weiguang Ye[1], Tianyi Han[1], Yuheng He[1], Yuan Cai[1], Ning Wang*

[1]*Department of Physics and the William Mong Institute of Nano Science and Technology, the Hong Kong University of Science and Technology, Hong Kong, China*

[2]*Department of Condensed Matter Physics, Group of Applied Physics, University of Geneva, 24 Quai Ernest Ansermet, CH1211 Geneva, Switzerland*

**Two-dimensional (2D) materials, such as graphene[1,2] and transition metal dichalcogenides[3-11] have attracted great attention because of the rich physics and potential applications in next-generation nano-sized electronic devices. Recently, atomically thin black phosphorus[12-15] (BP) has become a new member of the 2D materials family with high theoretical mobility[16] and tunable bandgap structure[17-19]. However, degradation of properties under atmospheric conditions[20,21] and high-density charge traps[14] in BP have largely limited its mobility ($\sim 400$ cm$^2$V$^{-1}$s$^{-1}$ at room temperature[12-15]) and thus restricted its future applications. Here, we report the fabrication of stable BN-BP-BN heterostructures by encapsulating atomically thin BP between hexagonal boron nitride (BN) layers to realize ultraclean BN-BP interfaces which allow a record-high field-effect mobility $\sim 1350$ cm$^2$V$^{-1}$s$^{-1}$ at room temperature and on-off ratios over $10^5$. At low temperatures, the mobility reaches $\sim 2700$ cm$^2$V$^{-1}$s$^{-1}$ and quantum oscillations in BP 2D hole gas are observed at low magnetic fields. Importantly, the BN-BP-BN heterostructure can effectively avoid the quality degradation of BP in ambient condition.**

Two-dimensional (2D) materials with both high mobility and high on-off ratios are required to expand applications in next-generation nanodevices. Graphene[1, 2], the most widely studied material, has shown rich physics and high mobility, while the absence of bandgap[1, 2] limits its further applications. Analogous to 2D transition metal dichalcogenides (TMDs) structures[3-11], atomically thin black phosphorus (BP)[12-15] has recently evoked interest due to its unique properties applicable in electronic and optoelectronic devices[22-24] since it has high theoretical mobility[16], tunable direct bandgap[17, 25] and ambipolarity[12-15]. Recently, the field-effect transistors[12-15, 26-28], electrically tunable PN junctions[23], radio-frequency transistors[29], and hetero-junctions[30, 31] based on few-layer phosphorene have also been demonstrated. However, the room-temperature mobility of few-layer phosphorene reported so far is limited to 400 $cm^2V^{-1}s^{-1}$ due to the presence of high-density charge traps and phonon scatterings[12-15]. The quality degradation of BP under the atmospheric conditions is mainly due to the reaction with $O_2$ saturated $H_2O$.[20, 21]

Here, we demonstrate a record-high room temperature field-effect mobility of about 1350 $cm^2V^{-1}s^{-1}$ and high on-off ratios over $10^5$ in few-layer phosphorene encapsulated by atomically thin hexagonal boron nitride (BN) which forms a stable BN-BP-BN heterostructure. The room-temperature mobility of ~1000 $cm^2V^{-1}s^{-1}$ has been rarely achieved in 2D semiconductor electron gas systems prepared by mechanical exfoliation techniques. In addition, the BN layer further avoids quality degradation of BP when exposed to atmosphere. At cryogenic temperatures, the mobility reaches ~2700 $cm^2V^{-1}s^{-1}$ and allows Shubnikov-de Haas (SdH) oscillations in BP at low magnetic fields (~6 T). Our fabrication and treatments for BN-BP-BN heterostructures open up a way to achieve high performance 2D semiconductors with largely improved room temperature mobility, on-off ratios, and stability under ambient conditions for practical applications in high-speed electronic and optoelectronic devices.

To achieve both high mobility and stability of BP field-effect transistors (FET) under atmospheric conditions, the sandwiched BN-BP-BN configuration and high-temperature annealing are the two key factors. The ultra-clean BN-BP interfaces are ensured by using the polymer-free van der Waals transfer technique[32] as shown in Fig. 1a. The few-layer BP mechanically exfoliated on a 300nm-$SiO_2$/Si substrate was first picked up by a thin BN flake (6-20nm). Then the BN-BP sample was transferred to a BN flake supported on a $SiO_2$/Si substrate to form the BN-BP-BN heterostructure. The atomically thin BP was completely encapsulated by two BN layers, which allows us to anneal the sample at temperatures up to 500 °C in argon atmosphere in order to further improve the sample quality. However, without using the BN protection layers, few-layer BP is easily broken down at 350 °C (Supplementary Fig. S3). In addition, the annealing process can significantly reduce the charge trap density in BP as no hysteresis effect is observed at room temperature (Fig. 2c and Supplementary Fig. S2).

Instead of using 1D edge-contact[32] which has been proven to be efficient for graphene encapsulated structure (but less efficient for 2D semiconductors), we developed the area contact technique for contacting the encapsulated 2D semiconductors. Beginning with the BN-BP-BN heterostructure, a hard mask is defined by standard electron-beam lithography technique using ZEP-520 resist (detailed process is shown in Supplementary Fig. S1). Since the $O_2$-plasma etching rates for BN and BP are different, the BN layer can be quickly etched while BP layer still survive as shown in Fig. 1c. At last, Cr/Au (2nm/60nm) electrodes are deposited using electron-beam lithography technique. Figs. 1b and c show the schematic and optical image of a BN-BP-BN Hall-bar device respectively. Note that no further annealing process is needed after deposition of the electrodes.

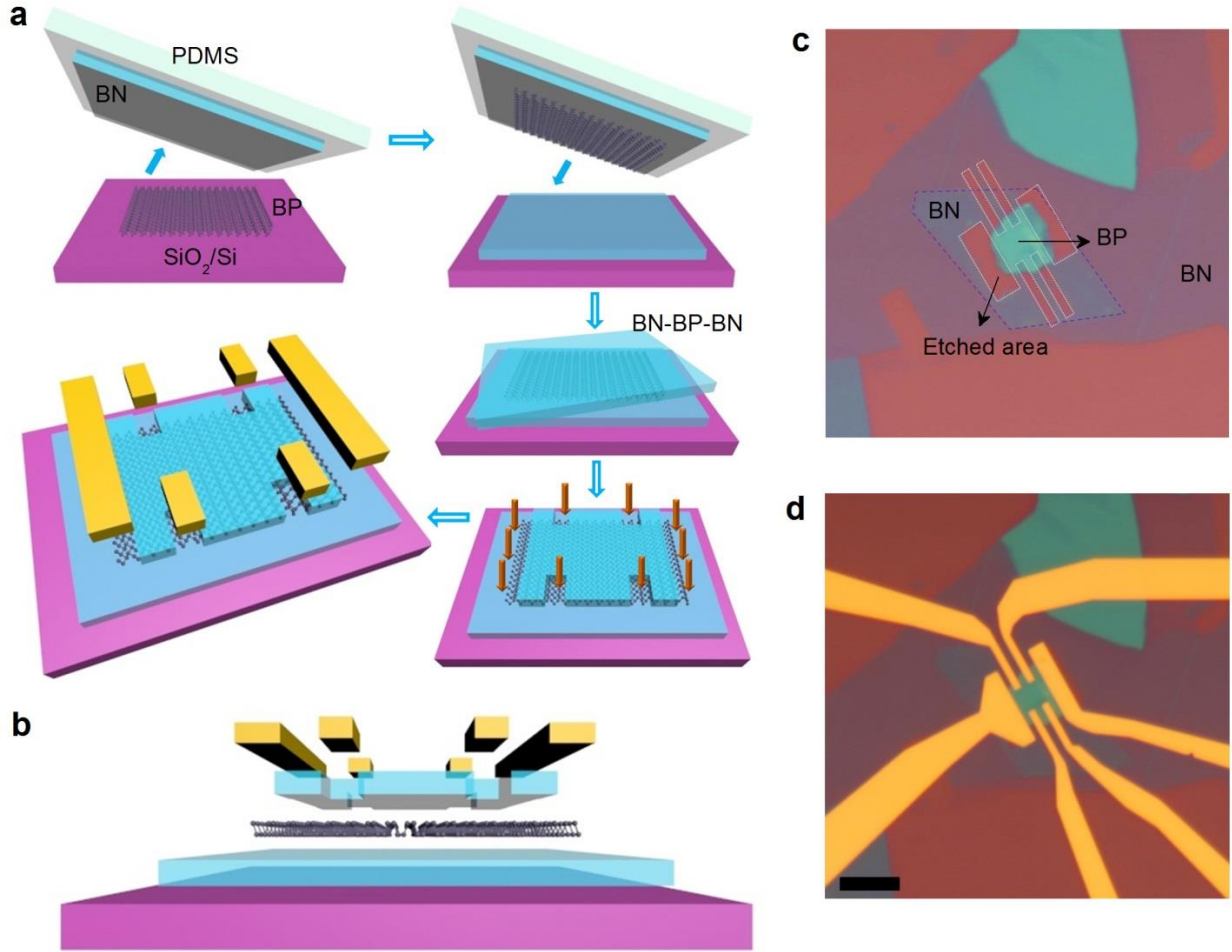

**Figure 1 | BN-BP-BN heterostructure device. a**, Schematic of the BN-BP-BN heterostructure device fabrication process. **b**,**d**, Schematic (**b**) and optical image (**d**) of a BN-BP-BN Hall-bar device. **c**, The BN-BP-BN heterostructure after $O_2$-plasma etching. The etched area is enclosed by white line. The purple-dashed line denotes the lower BN layer.

The high quality of the BN-BP-BN heterostructure is first confirmed by examining the field-effect mobility $\mu_F = \frac{1}{C_g}\frac{d\sigma}{dV_g}$ at different temperatures, where $\sigma$ is the conductivity and $C_g$ and $V_g$ are the gate capacitance and gate voltage respectively. Fig. 2a shows the conductivity as a function of gate voltage for an 8nm-thick BP sample. Similar to the results reported previously, both hole and electron conductance can be achieved (see the inset in Fig. 2a). Here we mainly focus on the hole conductance due to its higher mobility. With the information of gate capacitance $C_g = 1.1\,\text{F}\,\text{cm}^{-2}$ based on the thickness of SiO$_2$ (300 nm) and lower BN layer (~15nm determined by AFM), we extracted the hole $\mu_F$ from the linear part of conductance at different temperatures (Fig. 2b). As shown in Fig. 2a, we achieved high mobility $\mu_F \sim 1350\,\text{cm}^2\text{V}^{-1}\text{s}^{-1}$ in the 8nm-thick BN-BP-BN heterostructure at room temperature which is much larger than that observed in MoS$_2$[4, 6, 11] and BP on SiO$_2$[12-15]. Such a high quality of the BN-BP-BN heterostructure is further confirmed by the observation of a high on-off ratio of over $10^5$ and insignificant hysteresis at room temperature (Fig. 2c), while previous works[14] on BP supported on SiO$_2$ show a pronounced hysteresis ($\Delta V_g \sim 80$ V) depending on the sweep direction at room temperature, indicating the presence of a high density of charge traps. The small charge trap density in our BN-BP-BN heterostructure is not only due to the ultraclean BN-BP interfaces but also attributed to the high temperature annealing (300 °C - 500 °C) of the encapsulated BP. As shown in Fig. S2, in a BN-BP-BN heterostructure device without annealing, there exists a hysteresis $\Delta V_g \sim 10$ V (which it is much smaller than that observed in BP on SiO$_2$). These results confirm that high temperature annealing can effectively suppress charge trap states and further improve the mobility and on-off ratio of BP devices.

The stability of the electrical performance of the BN-BP-BN heterostructure is also examined at ambient conditions (with humidity around 60%). As shown in Fig. 2d, the mobility and on-off ratios show no degradation after one week exposure to ambient conditions. This shows that the BN-BP-BN heterostructure completely isolates BP layer from ambient conditions without reacting with O$_2$ saturated H$_2$O. The BN-BP-BN heterostructure with high room-temperature mobility, on-off ratios and stability make it a promising candidate for practical applications in electronic and optoelectronic devices.

At 1.7 K, the mobility of BN-BP-BN heterostructure reaches ~ 2700 cm$^2$V$^{-1}$s$^{-1}$ (Fig. 1a) and Hall mobility $\mu_H = \dfrac{\sigma}{n_h e}$ is ~ 1500 cm$^2$V$^{-1}$s$^{-1}$ at large gate voltage, where $n_h$ is the carrier density from Hall measurement, which is equal to the value extracted from the gate capacitance $n = C_g(V_g - V_{th})/e$ (Supplementary Fig. S4). The on-off ratios also exceed $10^8$ at 1.7 K. The temperature-dependence of the mobility in the BN-BP-BN heterostructures (Fig. 2b) shows a similar trend as reported previously for few-layer phosphorene[12] and monolayer MoS$_2$[4]. The mobility saturates at low temperatures (T<80 K), while it follows the expression $\mu \sim T^{-\gamma}$ at higher temperatures (T>80 K) due to phonon scatterings. The extracted value $\gamma$ is ~0.57 (at $V_g = -60$ V) for the 8-nm BP heterostructure and ~0.69 for the 15nm-thick (at $V_g = -50$ V) BP heterostructure. The obtained $\gamma$ in the BN-BP-BN heterostructures is consistent with the values observed in monolayer $MoS_2$ covered with high-κ dielectric[4] and few-layer phosphorene[12], but is much smaller than the values in bulk BP[33] and other 2D materials[34]. In the 15nm-thick BP heterostructure, we also obtained a high field-effect mobility $\mu_F \sim 850$ cm$^2$V$^{-1}$s$^{-1}$ at room temperature and ~1730 cm$^2$V$^{-1}$s$^{-1}$ at 1.7 K (Supplementary Fig. S5). Although $\mu_F$ is smaller than that in the 8nm-thick BP sample (~2700 cm$^2$V$^{-1}$s$^{-1}$), the Hall mobility reaches ~1490 cm$^2$V$^{-1}$s$^{-1}$ in the 15nm-thick BP sample at 1.7 K (Supplementary Fig. S5).

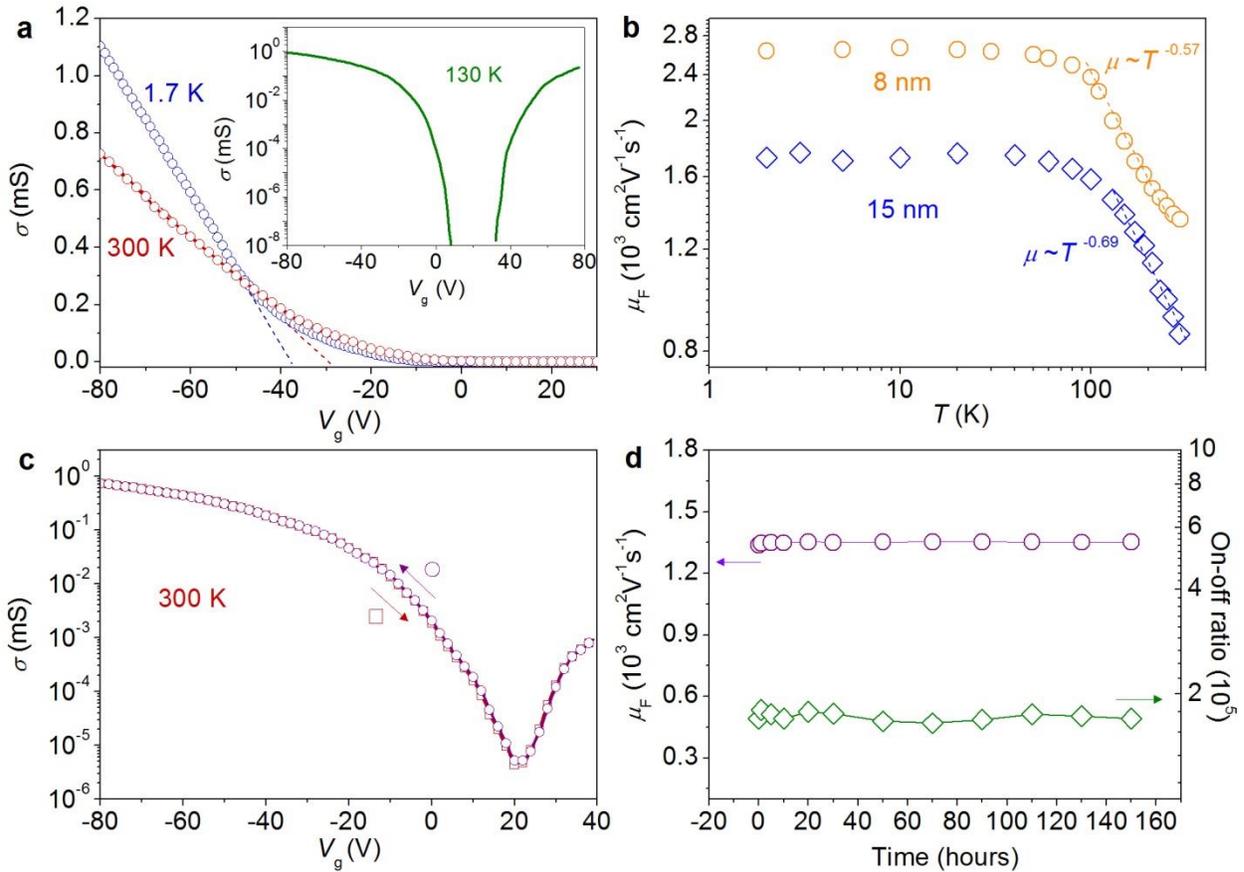

**Figure 2 | Mobility and stability of the BN-BP-BN heterostructure device. a**, Conductivity in the 8nm-thick BP heterostructure which shows a mobility ~1350 cm$^2$V$^{-1}$s$^{-1}$ at room temperature

and ~ 2700 $cm^2V^{-1}s^{-1}$ at 1.7 K. Inset shows the ambipolarity of the BP conductance. **b**, Temperature dependence of the field-effect mobility $\mu_F$ of the 8nm-thick (at $V_g = -60$ V) and 15nm-thick (at $V_g = -50$ V) BP heterostructures. Dashed lines serve as guidelines for the $\mu \sim T^{-\gamma}$ dependence. **c**, Conductivity at room temperatures show no hysteresis in the 8nm-thick BP heterostructure. **d**, Mobility and on-off ratio of the 8nm-thick BP heterostructure as function of ambient exposure time which shows no quality degradation after one week exposure.

The high quality of BN-BP-BN heterostructures has been reflected by the quantum oscillations observed at magnetic field $B > 6$ T. Resistance at different gate voltages shows a negative magneto-resistance (MR) (Fig. 3a inset), indicating a weak localization effect[35] previously observed in 2D electron gas systems containing disorder. In this kind of 2D electron gas systems, the time-reversal symmetry of closed electron path is broken by applying a magnetic field and back scatterings of electrons are suppressed resulting in a negative correction to the resistance. When $B > 6$ T (corresponding to a mobility $\sim 1/B$ around 1670 $cm^2V^{-1}s^{-1}$), we start to see SdH oscillations at higher gate voltages where the Hall mobility becomes high enough (Fig. 3a). The oscillation features are seen clearly in the form of $dR/dB$ as shown in Fig. 3b.

The SdH oscillations[36, 37] in 2D electron gas follow a $1/B_F$ period and the longitudinal resistance is given by $\Delta R = R(B,T)\cos[2\pi(B_F/B + 1/2 + \beta)]$, where $R(B,T)$ is the amplitude and $\beta$ ($0 \leq \beta \leq 1$) is the Berry phase. $\beta$ is known to be 0 in 2D electron gas and $1/2$ in graphene[1, 2]. To examine the 2D nature of hole gas in few-layer phosphorene, we plot $dR/dB$ as function of $1/B$ for different gate voltages. As shown in Fig. 3c, the oscillations at different gate voltages are periodic but with different oscillation frequencies. Smaller period $1/B_F$ is observed at higher gate voltages where carrier density $n$ is larger, which is consistent with the theoretical oscillation period[36] expressed as $1/B_F = 2/\Phi_0 n$, where $\Phi_0 = 4.14 \times 10^{-15}$ $Tm^2$ is the flux quantum. For example, at $V_g = -50$ V the oscillation period is ~ 0.0138 $T^{-1}$, corresponding to a carrier density of ~ $3.5 \times 10^{12}$ $cm^{-2}$ which is in excellent agreement with the value obtained by gate capacitance $n = C_g(V_g - V_{th})/e$ (Fig. 3e). Similar quantum oscillations are also observed in the 15nm-BP heterostructure as shown in Supplementary Fig. S6. Finally, Landau level (LL) index N is plotted and the linear fit yields a Berry phase $\beta = 0$ in our BP samples (Fig. 3d). Smaller LL index can be achieved if magnetic fields could be further increased.

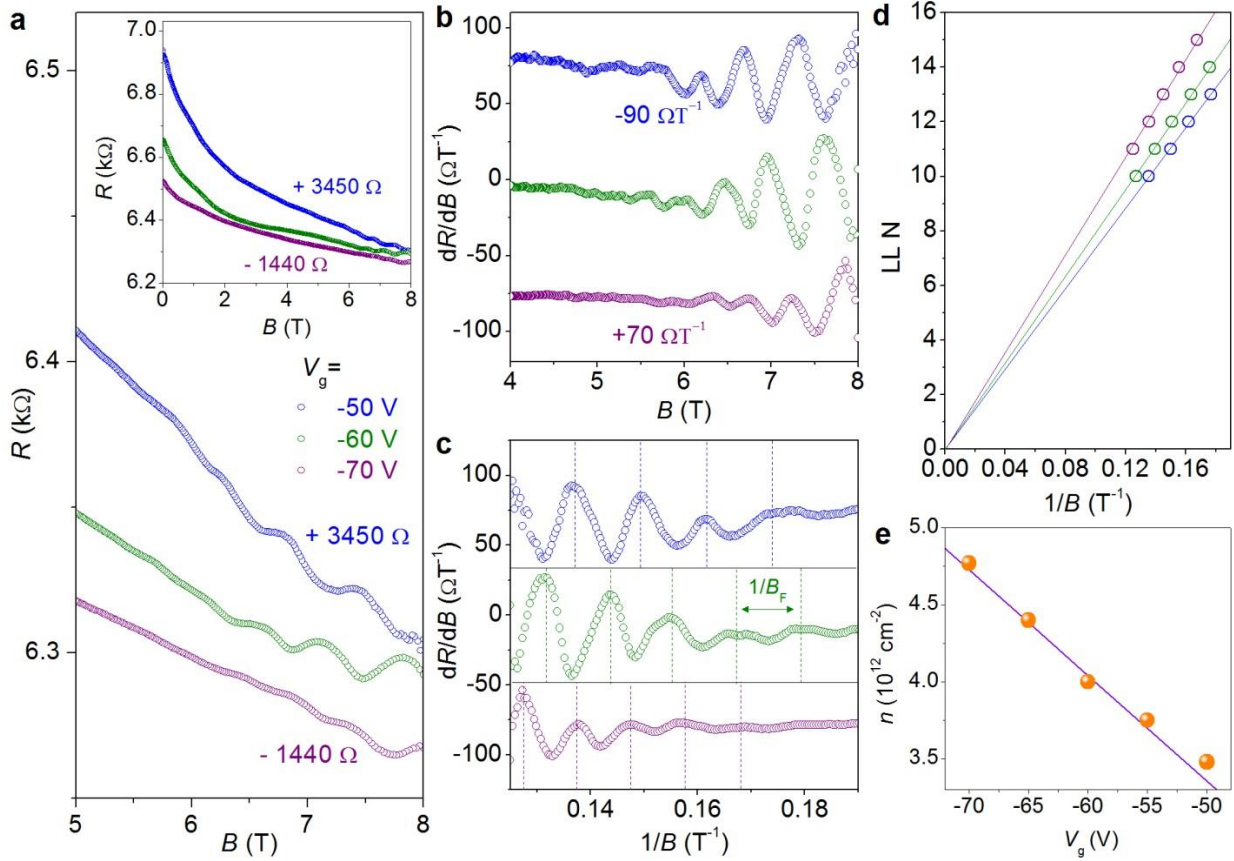

**Figure 3 | SdH oscillations in the 8nm-thick BN-BP-BN heterostructure. a**, Resistance changes plotted as function of magnetic fields at -50 V (blue dots), -60 V (green dots) and -70 V (purple dots) gate voltages respectively. The inset shows negative MR. **b**, The differential resistance $dR/dB$ plotted as function of magnetic field $B$ shows clearer SdH oscillations. **c**, $dR/dB$ plotted as function of $1/B$ yields an oscillation period $1/B_F$ which becomes smaller at higher gate voltages. **d**, Landau diagram at different gate voltages which yields a Berry phase $\beta = 0$. **e**, Carrier density at different gate voltages determined from fitting of LLs (orange dot) and gate capacitance (purple line).

In summary, we have demonstrated high-quality BN-BP-BN heterostructures achieved by encapsulating atomically thin BP between BN layers followed by annealing at high temperature.

The BN-BP-BN heterostructure protects BP from property degradation and allows us to investigate the intrinsic properties of BP. The BN-BP-BN samples show excellent stability in atmospheric environment with a high mobility ($\sim 1350$ cm$^2$V$^{-1}$s$^{-1}$) and on-off ratio over $10^5$. Quantum oscillations and zero Berry phase are observed in BP hole gas at a magnetic field of 6 T at cryogenic temperatures. The ultraclean interfaces realized by our fabrication process effectively suppress charge trap states and shed important light on improving the quality of atomically thin BP for practical applications in BP-based nanoelectronic devices.

## Methods

Bulk phosphorus and boron nitride crystals (Polartherm grade PT110) were purchased from Smart-elements and Momentive respectively. All atomically thin flakes are mechanically exfoliated based on the scotch-tape microcleavage method. Thickness of BN and BP flakes are determined by an atomic force microscope (Veeco-Innova). The BN-BP-BN heterostructure is fabricated using the polymer-free van der Waals transfer technique[32]. The encapsulated BP is annealed at an argon atmosphere at 300 °C - 500 °C for 8 hours before upper BN is etched. No further annealing process is needed after deposition of the electrodes. Electrical measurements are performed using lock-in techniques in a cryogenic system (1.7K-300 K and magnetic field up to 8T).

## Author information

[†]Xiaolong Chen and Yingying Wu contributed equally to this work.

## Corresponding Author

*E-mail: phwang@ust.hk.

## Acknowledgment

Financial support from the Research Grants Council of Hong Kong (Project Nos. HKU9/CRF/13G, 604112, HKUST9/CRF/08 and N_HKUST613/12) and technical support of the Raith-HKUST Nanotechnology Laboratory for the electron-beam lithography facility at MCPF (Project No. SEG_HKUST08) are hereby acknowledged.

**Author contributions**

N. W. and X. C. conceived the projects. X. C. and Y. Y. W. conducted most experiments including sample fabrication, data collection and analyses. N. W. is the principle investigator and coordinator of this project. X. C. and N. W. provided the physical interpretation and wrote the manuscript. Other authors provided technical assistance in sample preparation, data collection/analyses and experimental setup.

**Competing financial interests**

The authors declare no competing financial interests.

# Supplementary Information

# for

# High quality sandwiched black phosphorus heterostructure and its quantum oscillations

Xiaolong Chen[1,†], Yingying Wu[1,†], Zefei Wu[1], Shuigang Xu[1], Lin Wang[2], Yu Han[1], Weiguang Ye[1], Tianyi Han[1], Yuheng He[1], Yuan Cai[1], Ning Wang*

[1]*Department of Physics and the William Mong Institute of Nano Science and Technology, the Hong Kong University of Science and Technology, Hong Kong, China*

[2]*Department of Condensed Matter Physics, Group of Applied Physics, University of Geneva, 24 Quai Ernest Ansermet, CH1211 Geneva, Switzerland*

*Correspondence should be addressed to: Ning Wang, phwang@ust.hk

## 1. The process flow for making BN-BP-BN heterostructures

To obtain an ultraclean BN-BP interface, the polymer-free van der Waals transfer technique[1] is adopted to pick up BP flakes by BN. Detailed process for this transfer technique has been reported in Ref. 1. Therefore, we demonstrate here the processes after assembling and transferring BN-BP-BN heterostructures onto $SiO_2$/Si substrates as shown in Fig. S1a.

To make a hard mask, 600 nm e-beam resist ZEP layers are prepared and then standard e-beam lithography is used to pattern the ZEP layers as shown in Fig. 1b. The exposed area is etched by $O_2$-plasma. Since $O_2$-plasma etching rates for BN and BP are different, the BN layer can be quickly etched away while the BP layer still survives as shown in Fig. S1c. The white dashed line in Fig. S1d shows the etched area for electrode deposition after removing the hard mask. Fig. S1e shows the optical image of a Hall-bar device in BN-BP-BN configurations.

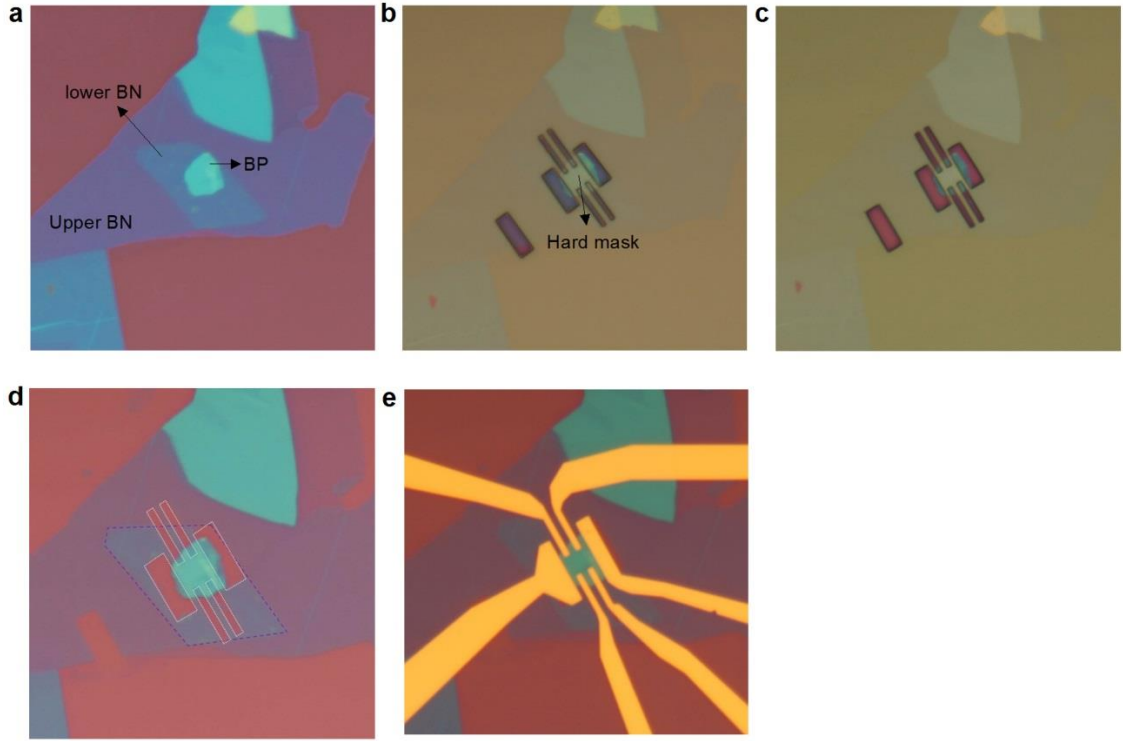

**Fig. S1 | The process flow for fabrication of the BN-BP-BN heterostructure device. a**, Optical image of the BN-BP-BN heterostructure on SiO$_2$. **b**, Defined area of the hard mask for etching of BN layers. **c**, BP layers remain after etching, while BN layers are etched away. **d**, The BN-BP-BN heterostructure after removing the ZEP mask. **e**, Optical image of a fabricated BN-BP-BN heterostructure Hall-bar device.

2. **High-temperature annealing of BP flakes**

High temperature annealing (300 ℃ - 500 ℃) of the entirely encapsulated BP flakes in argon (Ar) atmosphere before etching the upper BN layers can further improve the performance and quality of BP. As shown in Fig. 2c in the main text, the conductance of the 8nm-BP heterostructure shows no hysteresis at room temperature after annealing in Ar atmosphere at 350 ℃ for 8 hours. The 15nm-BP heterostructure shown in the main text is annealed at 400 ℃ for 8 hours and also shows no hysteresis at room temperature (Fig. S5b). The conductance of a 7.5nm-BP heterostructure without annealing treatment (Fig. S2) shows a hysteresis $\Delta V_\text{g}$ ~10 V at room temperature, although it is much smaller than the value observed in BP on SiO$_2$[2]. The observed hysteresis is due to the charge trapping effect (showing a positive direction of the hysteresis) instead of the capacitive coupling to BP as discussed in Ref. 3. Obviously, the high-temperature annealing for the encapsulated BP can effectively suppress the charge trapping effect and improve the performance of the BN-BP-BN heterostructures. However, without covering BN layers, thin BP layers prepared on SiO$_2$ were quickly destroyed by annealing at high temperatures. As shown in Fig. S3, three BP samples are destroyed after annealing at 350 °C for 8 hours.

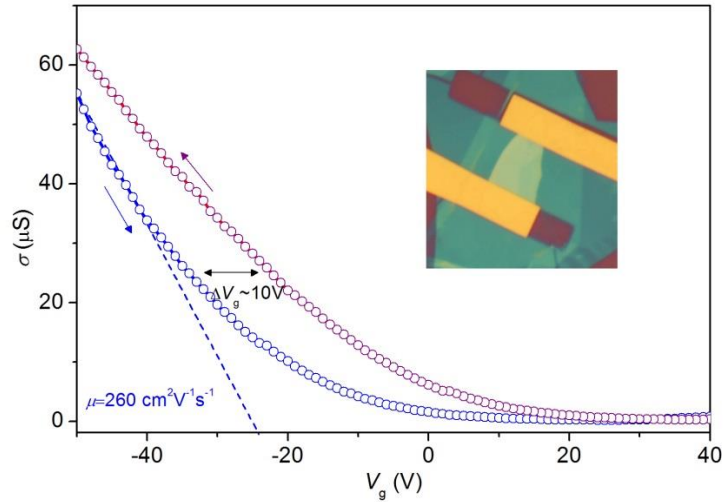

**Fig. S2 | The hysteresis effects in the BN-BP-BN heterostructure without annealing.** Without annealing treatment, the conductivity of a 7.5nm-BP heterostructure shows a hysteresis of about $\Delta V_g$ ~10 V at 300 K.

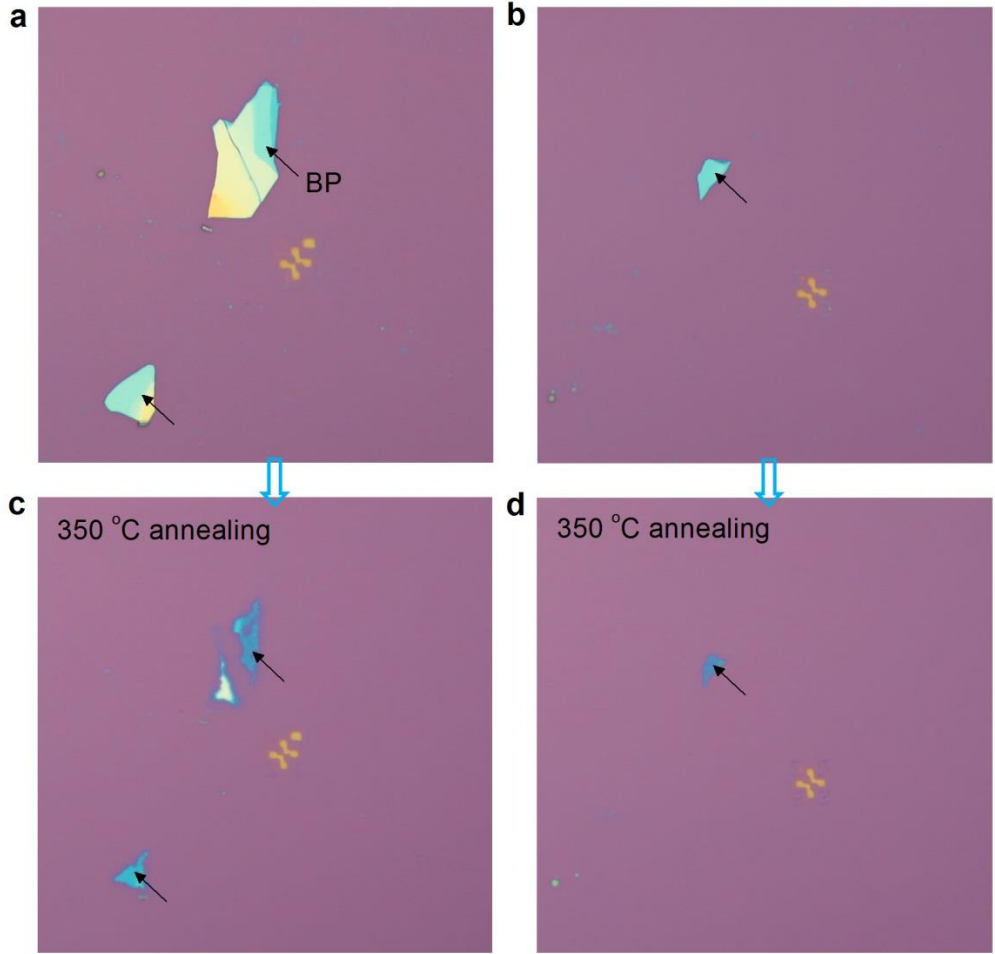

**Fig. S3 | High-temperature annealing of BP flakes on SiO$_2$. a-d**, Optical images of three BP flakes on a SiO$_2$ substrate before (**a**,**b**) and after annealing (**c**,**d**) at 350 °C for 8 hours in argon atmosphere.

## 3. Determination of carrier densities

The Hall mobility is calculated according to $\mu_H = \dfrac{\sigma}{n_h e}$, where $n_h = 1/R_H e$ is the carrier density determined from Hall measurements and $R_H = dR_{xy}/dB$ is the Hall coefficient. The extracted $n_h$ at different gate voltages (blue dot) is shown in Fig. S4. The linear fit of $n_h$ (red line) intercepts the x-axis at $V_g = -1$ V, which shows excellent agreement with the value obtained from the gate capacitance $n_g = C_g(V_g - V_{th})/e$ (blue dashed line), where $C_g = 1.1$ F cm$^{-2}$ and $V_{th} = -1$ V.

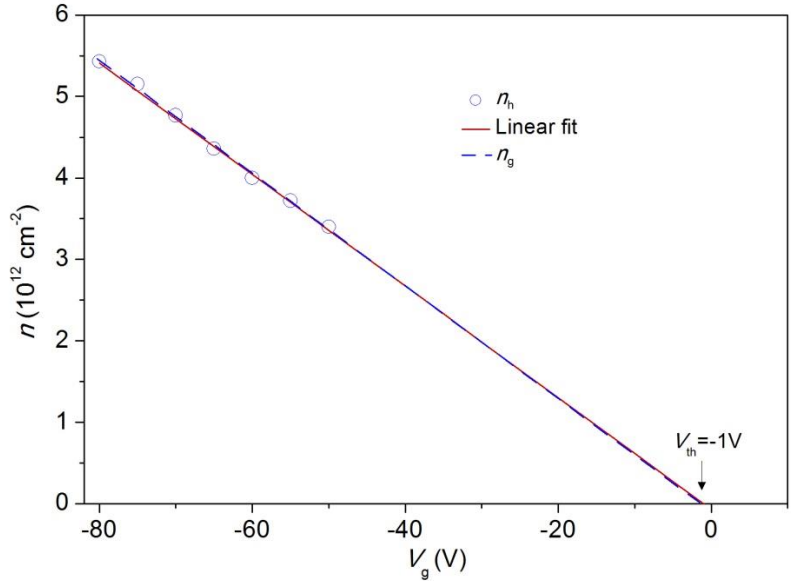

**Fig. S4 | Determination of carrier densities.** The carrier density is determined from Hall measurements (blue dot) and gate capacitance (blue dashed line) at 1.7 K. The red line is the linear fit of $n_h$.

### 4. Estimation of the mobility and hysteresis in the 15nm-BP heterostructure

At 1.7 K, the field effect mobility of the 15nm-BP heterostructure reaches $\sim 1730 \text{ cm}^2\text{V}^{-1}\text{s}^{-2}$ as shown in Fig. S5. The Hall mobility is $\sim 1490 \text{ cm}^2\text{V}^{-1}\text{s}^{-2}$ calculated by $\mu_H = \dfrac{\sigma}{n_h e}$ and $n_h$ is determined by the Hall measurement (the inset in Fig. S5a). No hysteresis is observed in the 15nm-BP heterostructure at room temperature as demonstrated by the conductivity with different gate sweeping directions (Fig. S5b).

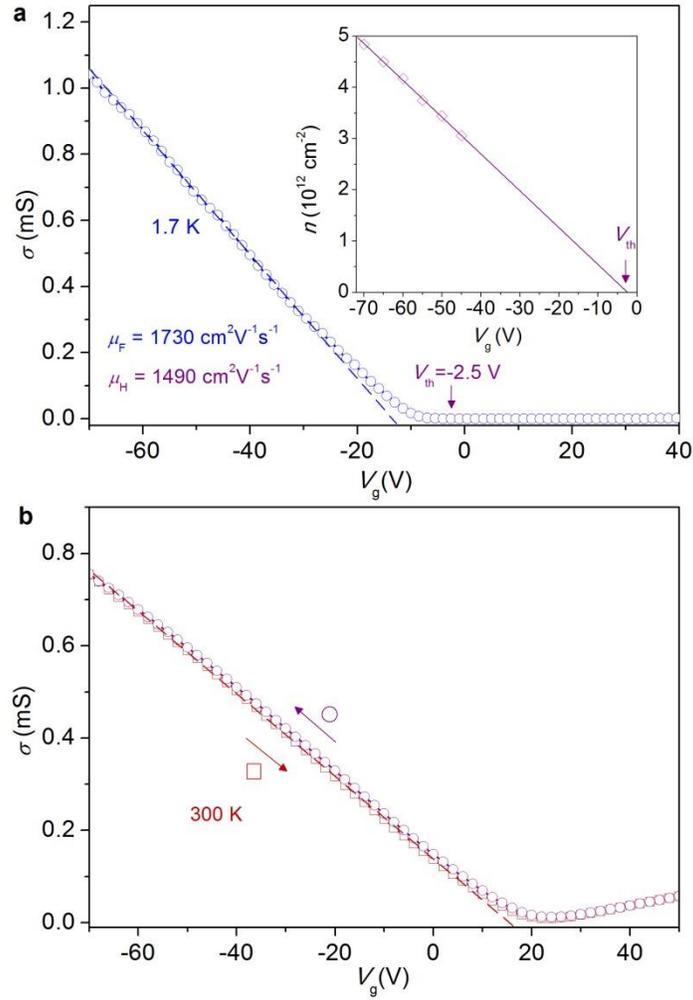

**Fig. S5 | Estimation of the mobility and hysteresis in the 15nm-BP heterostructure. a**, The conductivity at 1.7 K with a field effect mobility $\sim 1730\ \text{cm}^2\text{V}^{-1}\text{s}^{-2}$ and Hall mobility $\sim 1490\ \text{cm}^2\text{V}^{-1}\text{s}^{-2}$. The inset shows the carrier density determined from Hall measurements. **b**, The conductivity obtained under different gate sweeping directions, showing no hysteresis at 300 K.

## 5. SdH oscillations in the 15nm-BP heterostructure

The variation of measured resistance under different magnetic fields in the 15nm-BP heterostructure is consistent with that observed in the 8nm-BP sample including the negative MR (Fig. S6a), clear oscillation period of $1/B_F$ (Fig. S6b) and zero Berry phase (Fig. S6c).

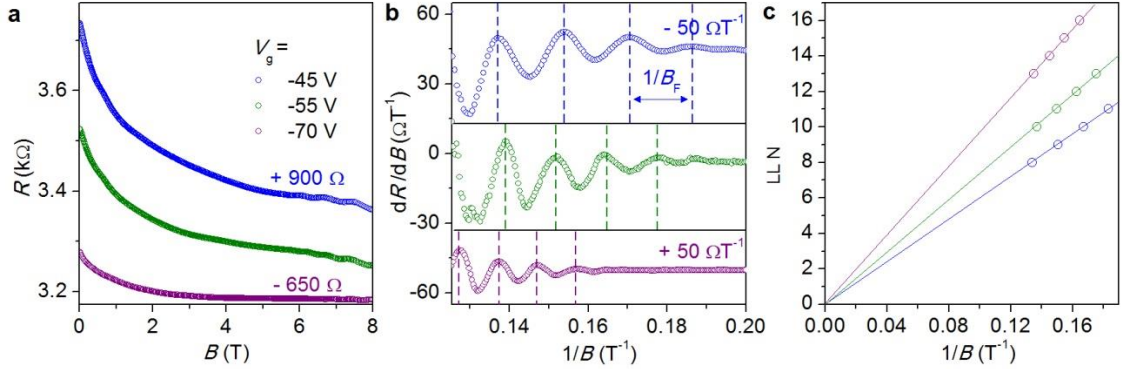

**Fig. S6 | SdH oscillations in the 15nm-BP heterostructure. a**, The measured resistance of the 15 nm-BP sample plotted as a function of magnetic fields at -45 V (blue dots), -55 V (green dots) and -70 V (purple dots) gate voltages respectively. **b**, $dR/dB$ plotted as a function of $1/B$ yields an oscillation period in $1/B_F$ which becomes smaller at a higher gate voltage. **c**, Landau diagram at different gate voltages.